\documentclass[a4paper,12pt]{article}
\usepackage{amssymb}

\usepackage{makeidx}
\usepackage{amsmath}
\usepackage{a4}

\input{tcilatex}

\begin{document}

\author{F. Bass, V. Freilikher, and O. Shefranova \and \textit{Department of
Physics, Bar-Ilan University,} \and \textit{\ Ramat Gan 52900, Israel}}
\title{Effect of space-time dispersion on the propagation of electromagnetic
waves in photonic crystals }
\maketitle

\begin{abstract}
We study the influence of the space and time dispersion on the frequency
dependence of the wave vectors of electromagnetic waves propagating in
three-dimensional photonic crystals. Two types of structures are considered:
media with weak periodic modulation of the permittivity, and photonic
crystals composed of the periodically arranged identical resonant dielectric
particles. It is shown that in these systems, in contrast to electrons in
solid crystals, different types of excitations exist. For example, a
peculiar kind of polaritons arises in the photonic crystals due to the
interaction of the electromagnetic field, eigenoscillations of the
dielectric medium, and Debye resonance. The widths of the transparency zones
and of the band gaps have been calculated as functions of the frequency and
of the parameters of the media. It is shown that in the photonic crystals
with dispersion, the number of transparency bands is larger than in
non-dispersive systems, and the width of the gaps in the frequency spectrum
of photons depends on the wave vector. The interaction of different types of
waves deforms the Brillouin zone, so that it may not have a plane boundary
(for example, a sphere), in which case the classical Bragg condition does
not hold.

\end{abstract}

\section{\protect\bigskip Introduction}

The propagation of waves in periodic structures is a topic of profound
importance, both for understanding the general properties of
matter-radiation interactions, and for the ever-increasing number of
practical applications (for example, photonic crystals) \cite{Yariv}, \cite%
{Jannopo}, \cite{Sakoda}. In spite of the close similarity to the
quantum-mechanical problem of electrons in a crystal lattice, the
propagation of classical waves in photonic crystals exhibits many rather
unusual distinctive features. The most dramatic example of such a difference
is the effect of time and space dispersion, which has no analogue in the
motion of conduction electrons. While the propagation of electromagnetic
waves in homogeneous dispersive media has been thoroughly studied \cite%
{Landau}, the effect of dispersion on the radiation propagating in periodic
structures still remains poorly understood. The exceptions are periodic
multiple quantum well structures (periodic arrays of dielectric layers), for
which the polariton spectrum and the optical response have been studied in
considerable detail \cite{wells1}, \cite{wells2}, \cite{wells3}, \cite%
{wells4}. In the recent publication \cite{pilozzi}, the exciton-polariton
susceptibility matrix was calculated, taking into account the well-well
interaction, and the polariton \ dispersion curves for an infinite periodic
array of quantum wells have been computed. \ Some interesting
polariton-induced effects were discovered. However, the one-dimensionality
of the system considered prevented the investigation of important effects
inherent only in dispersive media of higher dimensionality. In the present
paper, we deal with the linear wave equation of the most general type and
study the spectral properties of dispersive periodic dielectric media,
namely, the effect of the time and space dispersion on the frequency
dependence of the wave vector. The methods employed (small perturbation and
local perturbation methods) impose no restrictions on the dimensionality of
the system. As formulated, the problem is closely related to the calculation
of the energy of an electron as a function of the wave vector in periodic
crystals. However, in contrast to the later case, the solution of the
corresponding dispersion equation for electromagnetic waves can have several
branches, i.e. different types of waves can propagate. The interaction of
these waves is the crucial factor in the formation of the photonic spectrum,
and gives rise to rather unusual effects. In particular, it deforms the
Brillouin zone, so that it may not have a plane boundary (for example, a
sphere), in which case the classical Bragg condition is violated.

We consider the most general form of the wave equation

$\qquad \qquad \qquad \qquad \qquad \qquad \qquad $%
\begin{equation}
\hat{H}(\frac{\partial }{i\partial \vec{r}},\frac{\partial }{i\partial t})E+%
\hat{h}(\vec{r},\frac{\partial }{i\partial \vec{r}},\frac{\partial }{%
i\partial t})E=0,  \label{1}
\end{equation}%
where $\hat{H}$ and $\hat{h}$ \ are arbitrary functions of $\dfrac{\partial 
}{i\partial \vec{r}}$ and $\dfrac{\partial }{i\partial t},$ and $\hat{h}$ is
a periodic function of\ $\vec{r}$ with periods $\vec{r}_{n}=n_{1}\vec{r}%
_{1}+n_{2}\vec{r}_{2}+n_{3}\vec{r}_{3},$ where $\vec{r}_{1,}$ $\vec{r}_{2},%
\vec{r}_{3}$ define the unit cell of the photonic crystal, and\ n$_{1},$ n$%
_{2},$ n$_{3}$ are integers. In non-dispersive media, Eq. (1) becomes a
classical second-order wave equation; space-time dispersion corresponds to
higher-order derivatives with respect to the coordinates and the time in Eq.
(1).

According to the Bloch theorem, the solution of this equation has the form\
\ \ \ \ \ \ \ \ \ \ \ \ \ \ \ \ \ \ \ \ \ \ \ \ \ \ \ \ \ \ \ \ \ \ \ \ \ \
\ \ \ \ \ \ \ \ \ \ \ \ \ \ \ \ \ \ \ \ \ \ \ \ \ \ \ \ \ $\ $%
\begin{equation}
E{\large (}\vec{r},t)=E(\vec{r})e^{i\left( \vec{k}\vec{r}-\omega (\vec{k}%
)t\right) },  \label{2as}
\end{equation}%
where $E(\vec{r})$ is a periodic function with periods $\vec{r}_{n}$.

Our goal is to investigate the spectrum of waves propagating in the periodic
system, i.e. to find the dependences, $\omega _{l}(\vec{k})$, of the
frequency $\omega $ on the wave vector $\vec{k}$ (the subscript $\mathit{l}$
indicates different types of waves). The periodic functions $E(\vec{r})$ and 
$\hat{h}$($\vec{r},\frac{\partial }{\partial \vec{r}})$ can be expanded in
Fourier series,

\begin{equation}
E(\vec{r})=\sum_{n}E_{n}e^{i\vec{k}_{n}\vec{r}},  \label{3a}
\end{equation}%
\begin{equation}
\hat{h}(\vec{r},\frac{\partial }{i\partial \vec{r}},\omega
)=\sum_{n^{^{\prime }}}e^{i\vec{k}_{n%
{\acute{}}%
}\vec{r}}h_{n%
{\acute{}}%
}(\frac{\partial }{i\partial \vec{r}},\omega ),  \label{3b}
\end{equation}%
where $\vec{k}_{n%
{\acute{}}%
}$ are the reciprocal lattice vectors multiplied by $2\pi .$

Substituting \ Eq. $\left( 2\right) $ and $\left( 3\right) $ in Eq. $\left(
1\right) $ and setting the coefficients of each $e^{i\vec{\varkappa}_{m}\vec{%
r}}$ equal to zero, we obtain the following infinite system of equations,

\begin{equation}
H_{m}(\vec{\varkappa}_{m}{\normalsize ,}\omega )E_{m}+\sum_{n}h_{m-n}\left( 
\vec{\varkappa}_{n}{\normalsize ,}\omega \right) E_{n}=0  \label{4}
\end{equation}%
with $\vec{\varkappa}_{m}=\vec{k}+\vec{k}_{m}$

This system of equations has a nontrivial solutions, $E_{n}$, if its
determinant vanishes,

\begin{center}
\begin{equation}
D\left( \vec{\varkappa}_{m},\omega \right) =0  \label{5}
\end{equation}
\end{center}

Equation ($\ref{5})$ is the dispersion equation that determines the function 
$\omega (\vec{k})$. From Eq. ($\ref{2as}),$ it follows that $\omega (\vec{k}%
) $ is a periodic function of $\vec{k}$ with periods $\vec{k}_{n}$ \cite%
{Anselm}, \cite{Brillouin}. To avoid problems associated with the infinite
rank of the determinant in Eq. (\ref{5}), we consider two limiting cases: a
weak periodic modulation that corresponds to the nearly-free electron
approximation,{\Large \ }and a crystal, for which the method of local
perturbation is applicable (an analog of the tight-binding approximation).

\section{Weak periodic modulation}

In this section, we assume that the amplitude of the periodic modulation of
the medium is small, so that

\begin{equation}
\left| \hat{h}\left( \vec{r},\frac{\partial }{i\partial \vec{r}},\frac{%
\partial }{i\partial t}\right) E\right| \ll \left| \hat{H}\left( \frac{%
\partial }{i\partial \vec{r}},\frac{\partial }{i\partial t}\right) E\right| ,
\label{6}
\end{equation}%
and the term $\hat{h}(r,\frac{\partial }{i\partial \vec{r}},\frac{\partial }{%
i\partial t})E$ in Eq. (\ref{1}) can be considered as a small perturbation.

When $\hat{h}=0$ (zero-order approximation)\ all $\ $coefficients $h_{m-n}$
in Eq$.$ (\ref{4}) are zero, $E_{m}=$ $0,$ except for $m=0$ ($E_{0}=const),$
and the dispersion equation for the homogeneous medium takes the form

\begin{center}
\begin{equation}
H\left( \vec{k},\omega \right) =0  \label{7}
\end{equation}
\end{center}

Note that only real solutions $\omega _{l}(\vec{k})$ correspond to
propagating waves. We denote the number of such waves by $L$. To find
coefficients $E_{n}$ to first order of the small $\left| \hat{h}\right| ,$
there is no need to deal with the exact dispersion equation, Eq. (\ref{5}).
Instead, we set to zero all $E_{n}$ with $n\neq 0$ in the second term in Eq.
(\ref{4}) and obtain

$\qquad \qquad \qquad \qquad \qquad \qquad \ \ \ \ \ \ $%
\begin{equation}
\ \ \ \ \ E_{n}=-\frac{h_{n}\left( \vec{k}_{n},\omega \right) }{H\left( \vec{%
k}_{n},\omega \right) }E_{0};\ \ \ \ \   \label{8b}
\end{equation}

$\qquad \qquad \ \ \ \qquad \qquad \qquad \qquad $%
\begin{equation}
\delta \omega _{l}=\frac{1}{\frac{dH}{d\omega }(\vec{k},\omega _{l})}\sum_{n}%
\frac{[h_{n}(k_{n,}\omega _{l})]^{2}}{H(k_{n},\omega _{l})}.  \label{8d}
\end{equation}

It follows from Eq. (\ref{6}) that $\left| E_{m}\right| <<\left|
E_{0}\right| $ except for the case when there exists some $m=\nu $ for which 
$H(\vec{k}_{\nu },\omega (\vec{k}_{\nu }))$ is small. In this instance $%
E_{\nu }$ in Eq. (\ref{4}) is not small compare to $E_{0},$ and one has to
keep in Eq. (\ref{4}) also terms containing $E_{\nu }.$ This yields\bigskip 
\begin{eqnarray}
H\left( \vec{k},\omega \right) E_{0}+h_{-\nu }\left( \vec{\varkappa}_{-\nu
},\omega \right) E_{\nu } &=&0  \label{9} \\
H\left( \vec{k},\omega \right) E\nu +h_{\nu }\left( \vec{\varkappa}_{\nu
},\omega \right) E_{0} &=&0  \label{9a}
\end{eqnarray}

The condition for nontrivial solutions of Eqs. (\ref{9}), (\ref{9a}) to
exist leads to the following dispersion equation in the periodically
modulated medium in the degenerate case

\begin{center}
\begin{equation}
H\left( \vec{k},\omega \right) H\left( \vec{\varkappa}_{\nu },\omega \right)
-h_{\nu }\left( \vec{k},\omega \right) h_{-\nu }\left( \vec{\varkappa}_{-\nu
},\omega \right) =0.  \label{10}
\end{equation}
\end{center}

To solve the dispersion equation Eq. (\ref{10}) we\ write $H\left( \vec{k}%
,\omega \right) $in the form

\begin{equation}
H\left( \vec{k},\omega \right) =f\left( \vec{k},\omega \right)
\prod_{l=1}^{L}\left[ \omega -\omega _{l}\left( \vec{k}\right) \right] ,
\label{11a}
\end{equation}%
$\ $where function $f$ has no poles on the real axis. Substituting Eq. (\ref%
{11a}) into Eq. (\ref{10}), we obtain

\begin{equation}
\prod_{l=1}^{L}\left[ \omega -\omega _{l}\left( \vec{k}\right) \right]
\prod_{m=1}^{L}\left[ \omega -\omega _{m}\left( \vec{\varkappa}_{\nu
}\right) \right] -\frac{h_{\nu }\left( \vec{k},\omega \right) h_{-\nu
}\left( \vec{k},\omega \right) }{f\left( \vec{k},\omega \right) f\left( \vec{%
\varkappa}_{\nu },\omega \right) }=0.  \label{12}
\end{equation}%
\ If we write the solutions of Eq. (\ref{12}) as 
\begin{equation}
\omega \left( \vec{k}\right) =\omega _{l}\left( \vec{k}\right) +\delta
\omega _{l}\left( \vec{k}\right) ,\text{\ \ \ \ }(l=1,....L)  \label{omega}
\end{equation}%
and assume that for all non-interacting waves 
\begin{equation}
\left| \omega _{l}\left( \vec{k}\right) -\omega _{i}\left( \vec{k}_{\nu
}\right) \right| >>\delta \omega _{l}\left( \vec{k}\right) ,  \label{har}
\end{equation}%
the corrections $\delta \omega _{l}\left( \vec{k}\right) $can be found by
successive approximations. This procedure yields:%
\begin{equation}
\delta \omega _{l}=\frac{\left| h_{\nu }\left( \vec{k}_{\nu },\omega
_{l}\left( \vec{k}\right) \right) \right| ^{2}}{f\left( \vec{k},\omega
_{l}\left( \vec{k}\right) \right) f\left( \vec{k}_{\nu },\omega _{l}\left( 
\vec{k}_{\nu }\right) \right) \prod_{m\neq l}^{L}\left[ \omega _{l}\left( 
\vec{k}\right) -\omega _{m}\left( \vec{k}\right) \right] \prod_{m\neq l}^{L}%
\left[ \omega _{l}\left( \vec{k}\right) -\omega _{m}\left( \vec{k}_{\nu
}\right) \right] }  \label{delttt}
\end{equation}

It is important to note that this correction is of second order in the small
perturbation $h$.

The situation is different when for some $l=n,$ there exists $\vec{k}$ for
which $\omega _{n}(\vec{k})$ is close to $\omega _{n}(\vec{k}_{\nu }),$ and
inequality Eq. (\ref{har}) is violated. Then, the following approximate
equation for the dispersion curves $\omega _{n}(\vec{k})$ can be derived
from Eq. (\ref{12}):

\begin{equation}
\left[ \omega -\omega _{n}\left( \vec{k}\right) \right] \left[ \omega
-\omega _{n}\left( \vec{\varkappa}_{\nu }\right) \right] -\eta
_{n}^{2}\left( \vec{k},\vec{\varkappa}_{\nu }\right) =0  \label{14}
\end{equation}

\begin{equation}
\eta _{n}^{2}\left( \vec{k},\vec{\varkappa}_{\nu }\right) =\frac{\left|
h_{\nu }\left( \vec{k}_{\nu },\omega _{n}\left( \vec{k}\right) \right)
\right| ^{2}}{f_{n}\left( \vec{k},\omega _{n}\left( \vec{k}\right) \right)
f_{n}\left( \vec{\varkappa}_{\nu },\omega _{n}\left( \vec{\varkappa}_{\nu
}\right) \right) \prod_{l\neq 1}^{L}\left[ \omega _{n}-\omega _{l}\left( 
\vec{\varkappa}_{\nu }\right) \right] \prod_{l=1}^{L}\left[ \omega
_{n}-\omega _{l}\left( \vec{k}\right) \right] }  \label{15}
\end{equation}

The two solutions of the quadratic equation Eq. (\ref{14}) are:

\begin{equation}
\omega _{n\pm }\left( \vec{k}\right) =\frac{\omega _{n}\left( \vec{k}_{\nu
}\right) +\omega _{n}\left( \vec{k}\right) }{2}\pm \sqrt{\frac{\left( \omega
_{n}\left( \vec{k}\right) -\omega _{n}\left( \vec{\varkappa}_{\nu }\right)
\right) ^{2}}{4}+\eta _{n}^{2}\left( \vec{k},\vec{k}_{\nu }\right) }
\label{16}
\end{equation}

Note that for propagating waves, $\eta _{n}$ is real, i.e. $\eta _{n}^{2}>0.$
If

\begin{equation}
\omega _{n}\left( \vec{k}\right) =\omega _{n}\left( \vec{k}_{\nu }\right) ,
\label{16a}
\end{equation}%
then obviously%
\begin{equation}
\omega _{n\pm }\left( \vec{k}\right) =\omega _{n}\left( \vec{k}\right) \pm
\eta _{n}\left( \vec{k},\vec{k}_{\nu }\right) ,  \label{16b}
\end{equation}%
which means that there is a gap, $\Delta \omega _{n},$ in the frequency
spectrum

\begin{equation}
\Delta \omega _{n}\left( \vec{k},\vec{k}_{\nu }\right) =\omega _{n+}\left( 
\vec{k},\vec{k}_{\nu }\right) -\omega _{n-}\left( \vec{k},\vec{k}_{\nu
}\right) =2\eta _{n}\left( \vec{k},\vec{k}_{\nu }\right)  \label{17}
\end{equation}%
A similar gap exists in the energy spectrum of electrons in a periodic
potential \cite{Anselm}, with the following important difference: the width
of the gap in the frequency spectrum of photons, Eq. (\ref{17}), depends on
the wave vector $\vec{k}.$

Equation (\ref{16a}) defines the surface that is the boundary of the
Brillouin zone of the photon. For an isotropic medium, the frequency $\omega
_{n}$ depends on the modulus of the wave vector, and the solution of Eq. (%
\ref{16a}) is $k^{2}=k_{\nu }^{2}$. Hence, the equation for the Brillouin
zone boundary takes the form

\begin{equation}
2\vec{k}\vec{k}_{\nu }=-\vec{k}_{\nu }^{2},  \label{18}
\end{equation}%
which is the equation of a plane in $\vec{k}$-space. Eq. (\ref{18})
coincides with the well-known Bragg condition.

Much more unusual is the photonic spectrum when two roots of Eq. (\ref{10})
that belong to different branches (say, $\omega _{n}\left( \vec{k}\right) $
and $\omega _{q}\left( \vec{k}\right) $) are close to each other, i.e. $%
\omega _{n}\left( \vec{k}\right) \simeq $ $\omega _{q}\left( \vec{k}_{\nu
}\right) .$ Then, the solution of the dispersion equation Eq. (\ref{10}) is
given by Eq. (\ref{16}) with $\omega _{n}\left( \vec{k}_{\nu }\right) $
replaced by $\omega _{q}\left( \vec{k}_{\nu }\right) $ and $\eta _{n}\left( 
\vec{k},\vec{k}_{\nu }\right) $ substituted by

\begin{center}
\begin{equation}
\eta _{nq}^{2}\left( \vec{k},\vec{k}_{\nu }\right) =\frac{\left| h_{\nu
}\left( \vec{k}_{\nu },\omega _{n}\left( \vec{k}\right) \right) \right| ^{2}%
}{f_{1}\left( \vec{k},\omega _{n}\left( \vec{k}\right) \right) f_{2}\left( 
\vec{k}_{\nu },\omega _{n}\left( \vec{k}\right) \right) \prod_{l\neq n}^{L}%
\left[ \omega _{l}\left( \vec{k}\right) -\omega _{n}\right] \prod_{l\neq
q}^{L}\left[ \omega _{l}\left( \vec{k}_{\nu }\right) -\omega _{q}\right] }
\label{19}
\end{equation}
\end{center}

To exhibit the boundary of the Brillouin zone, we consider small values of
the wave vector, so that

\begin{eqnarray}
\omega _{i}\left( \vec{k}\right) &=&\omega _{i0}+\beta _{i}k^{2};\text{ \ \
\ \ }i=n,q  \label{20} \\
\beta _{i} &=&\frac{\partial \omega _{i}(0)}{\partial (k^{2})}
\end{eqnarray}%
The equality $\omega _{n}\left( \vec{k}\right) =\omega _{q}\left( \vec{k}%
_{\nu }\right) $ yields the following equation for the desired surface in $%
\vec{k}-$space:

\begin{equation}
\left( \vec{k}-\vec{k}_{R}\right) ^{2}=R^{2},  \label{21}
\end{equation}%
with 
\begin{equation*}
\vec{k}_{R}=\frac{\beta _{1}}{\beta _{2}-\beta _{1}}\vec{k}_{\nu }
\end{equation*}%
\begin{equation*}
R^{2}=\frac{\beta _{1}\beta _{2}}{\left( \beta _{2}-\beta _{1}\right) ^{2}}%
k_{\nu }^{2}+\frac{\omega _{2}(0)-\omega _{1}(0)}{\beta _{2}-\beta _{1}}
\end{equation*}%
Equation (\ref{21}) is the equations of a sphere with the center at $\vec{k}%
_{R}$ and having radius $R$. The spherical shape of the Brillouin zone
originates from the interaction of different types of waves ($n$ and $q)$,
and has no precedent in the theory of electrons in periodic lattices. In the
absence of interactions, Eq. (\ref{21}) loses its physical meaning, and $%
R^{2}$ becomes negative

It is interesting to point out another distinctive feature of photonic
periodic structures. In contrast to electrons, in electrodynamics the
radiation (or scattering) problem exists along with the eigenvalues problem
considered above.\ In this case, the frequency, $\omega $, is an external
parameter, and the dependence of the wave vector on the frequency, $%
\overrightarrow{k}(\omega ),$ is to be found. This dependence can be
obtained from the solution of the eigenvalues problem, $\omega (%
\overrightarrow{k})$. From Eqs. (\ref{omega}) and (\ref{16b}) it follows that

\begin{equation}
\overrightarrow{k}(\omega )=\overrightarrow{k}_{0}(\omega )+\delta 
\overrightarrow{k}(\omega ).  \label{ka}
\end{equation}%
When $\left| \delta \overrightarrow{k}\right| <<\left| \overrightarrow{k_{0}}%
\right| $%
\begin{equation}
\delta \omega =\overrightarrow{v}_{g}\delta \overrightarrow{k},  \label{veg}
\end{equation}%
where $\overrightarrow{v}_{g}=\frac{\partial \omega _{0}}{\partial 
\overrightarrow{k}}$ is the group velocity and $\theta $ is the angle
between $\overrightarrow{v}_{g}$ and $\overrightarrow{k_{0.}}$

\section{Periodic array of resonant particles}

In this section, we consider a photonic crystal composed of periodically
arranged identical dielectric particles having refractive index $n$, and
study the spectrum of excitations (eigenwaves) associated with a resonance
in a single particle. Such a resonance exists when the particle size, $d,$
is of order of the wavelength inside the particle, $\lambda =\frac{2\pi c}{%
n\omega },$ (Mi resonances, for example). This system is an analog of the
tight-binding model in the solid state theory \cite{Agranovich}, and
presents an opposite limiting case to that considered in the previous
section. Indeed, while in a medium with weak periodic modulation, the band
gap is small compared to the conduction band, the gaps in the spectrum of
the photonic crystal studied in this section are much wider than the passing
zones.

In what follows, we consider the frequency range in which the wavelength of
the radiation in the medium (having refractive index $n_{0})$ is larger than
the typical size of the particles 
\begin{equation}
d\ll \frac{2\pi c}{n_{0}\omega }  \label{22}
\end{equation}%
In this case, the problem can be handled by means of the local perturbation
method (LPM) \cite{fermi}, \cite{Lifsh}, \cite{fix}, \cite{freiliv}which is
based on the assumption that the (unknown) field is independent of the
coordinates inside each particle. In contrast to the Born approximation,
which considers only weak scattering, LPM leaves room for arbitrarily large
amplitudes of scattered fields, and does not rule out the existence of
resonances that take place when $d\approx \frac{2\pi c}{n\omega }$.

Equation (\ref{1}) now takes the form

\begin{equation}
\hat{H}(\frac{\partial }{i\partial \vec{r}},\frac{\partial }{i\partial t}%
)E+\sum_{p=-\infty }^{\infty }\widehat{U}\left( \vec{r}-\vec{r}_{p},\frac{%
\partial }{i\partial t}\right) E=0,  \label{23}
\end{equation}%
where the spatial-time dispersion of the homogeneous host medium is
incorporated in $\hat{H}$. The operator $\widehat{U}$ is non-zero only
inside the particles and depends on their shape, position, and dielectric
properties. For the sake of simplicity, we shall assume that the dielectric
particles possess only time dispersion.

Since $\widehat{U}$ in Eq. (\ref{23}) is a periodic function of $\vec{r}$,
the solution of Eq. (\ref{23}) can be written in the form of Eq. (\ref{2as}%
). If the inequality (\ref{22}) holds, the LPM approximation is applicable,
implying that 
\begin{equation}
U\left( \vec{r}-\vec{r}_{p},\omega \right) E\left( \vec{r}\right) e^{i\vec{k}%
\vec{r}}\approx U\left( \vec{r}-\vec{r}_{p},\omega \right) E\left( \vec{r}%
_{n}\right) e^{i\vec{k}_{p}\vec{r}_{p}}  \label{27}
\end{equation}%
where $E\left( \vec{r}\right) $ is a periodic function. If we assume that
one of the particles is located at the origin, $\vec{r}=0$, then $E\left( 
\vec{r}_{p}\right) =E\left( 0\right) .$ Substitution of Eq. (\ref{27}) into
Eq. (\ref{23}) yields:

\begin{equation}
\hat{H}(\frac{\partial }{\partial r},\omega )E\left( \vec{r}\right) +E\left(
0\right) \sum_{p=-\infty }^{\infty }U\left( \vec{r}-\vec{r}_{p},\omega
\right) e^{i\vec{k}_{p}\vec{r}}=0.  \label{26}
\end{equation}%
Equation (\ref{26}) can be rewritten in integral form,

\begin{equation}
E\left( \vec{r}\right) +E\left( 0\right) \sum_{p}\int G\left( \vec{r}-\vec{r}%
{\acute{}}%
\right) U\left( \vec{r}%
{\acute{}}%
-\vec{r}_{p},\omega \right) d\vec{r}%
{\acute{}}%
=0,  \label{28a}
\end{equation}%
where the Green function of the homogeneous ($U=0)$ host medium, $G,$ is
defined by the equation

\begin{equation}
\hat{H}(\frac{\partial }{\partial r},\omega )G\left( \vec{r},\vec{r}%
{\acute{}}%
\right) =\delta \left( \vec{r}-\vec{r}%
{\acute{}}%
\right)  \label{29}
\end{equation}%
Changing Eq. (\ref{29}) to spatial Fourier transforms, we obtain

\begin{center}
\begin{equation}
G\left( \vec{r},\vec{r}%
{\acute{}}%
\right) =\frac{1}{\left( 2\pi \right) ^{3}}\int \frac{e^{i\overrightarrow{k}%
\left( \vec{r}-\vec{r}%
{\acute{}}%
\right) }}{H(\overrightarrow{k},\omega )}d\overrightarrow{k}  \label{30}
\end{equation}
\end{center}

In an isotropic medium, $H(\overrightarrow{k},\omega )=H(k^{2},\omega ),$
the integral in Eq. (\ref{30}) can be calculated explicitly:

\begin{equation}
G\left( \vec{r}-\vec{r}%
{\acute{}}%
\right) =\frac{1}{4\pi \left| \vec{r}-\vec{r}^{%
{\acute{}}%
}\right| }\sum_{l=1}^{L}\frac{e^{-\left| \zeta _{l}\right| \left| \vec{r}-%
\vec{r}^{%
{\acute{}}%
}\right| }}{H%
{\acute{}}%
(-\zeta _{l}^{2},\omega )},  \label{32}
\end{equation}%
where $\zeta _{l}^{2}$ are roots of the equation

\begin{equation}
H(-\zeta _{l}^{2},\omega )=0,  \label{aaaaa}
\end{equation}%
and $H%
{\acute{}}%
=\frac{dH}{d(-\zeta _{l}^{2})}$.

By substituting $\vec{r}=0$ in Eq. (\ref{28a}) we obtain the following
dispersion equation for $\omega $ ($\vec{k}$)

\begin{equation}
1+\int G\left( \vec{r}%
{\acute{}}%
\right) U\left( \vec{r}%
{\acute{}}%
,\omega \right) d\vec{r}^{\prime }+\int U\left( \vec{r}%
{\acute{}}%
,\omega \right) d\vec{r}^{\prime }\sum_{n\neq 0}G\left( \vec{r}_{n}\right)
e^{i\vec{k}\vec{r}_{n}}=0  \label{33}
\end{equation}%
It is apparent that the second term in Eq. (\ref{33}) corresponds to the
resonance frequencies of an isolated single particle, while the third term
gives the corrections due to the multiple scattering (interactions). This
term is small compared to the second one, therefore, we seek the solutions
of Eq. (\ref{33}) of the form 
\begin{equation}
\omega _{p}=\omega _{0p}+\delta \omega _{p},  \label{34}
\end{equation}%
where $\omega _{0p}$ are the solutions of the ''zero-order'' (without
interactions) equation

\begin{equation}
1+\int G\left( \vec{r},\omega \right) U\left( \vec{r}%
{\acute{}}%
,\omega \right) d\vec{r}%
{\acute{}}%
=0,  \label{35}
\end{equation}%
and $\delta \omega _{p}<<\omega _{0p}.$

Since the summands in Eq. (\ref{32}) decrease exponentially with increasing $%
\left| \vec{r}_{n}\right| $, we may keep only the term with the maximal $%
\zeta _{l}$ (denote it as $\zeta _{m}).$ Then, for $\delta \omega _{p},$ we
find 
\begin{equation}
\delta \omega _{p}=\left\{ -\frac{\int U\left( \vec{r}%
{\acute{}}%
\right) d\vec{r}%
{\acute{}}%
}{\frac{d}{d\omega }H%
{\acute{}}%
(-\zeta _{m}^{2})\int G\left( \vec{r}%
{\acute{}}%
,\omega \right) U\left( \vec{r}%
{\acute{}}%
,\omega \right) d\vec{r}%
{\acute{}}%
}\sum_{n\neq 0}\frac{\exp (-\zeta _{m}\left| \vec{r}_{n}\right| )}{\left| 
\vec{r}_{n}\right| }e^{i\vec{k}\vec{r}_{n}}\right\} _{\omega =\omega _{0p}}
\label{36}
\end{equation}

For the same reason, in the sum in Eq. (\ref{36}), we take into account only
the terms corresponding to the interaction with nearest neighbors. For a
cubic crystal having side $b,$ this yields

\begin{eqnarray}
\delta \omega _{p} &=&-A(\cos k_{x}b+\cos k_{y}b+\cos k_{z}b),  \label{37} \\
A &=&-\frac{\int U\left( \vec{r}%
{\acute{}}%
\right) d\vec{r}%
{\acute{}}%
e^{-\left| \zeta _{m}\right| b}}{4\pi b\frac{d}{d\omega }H^{\prime }\left(
-\zeta _{m}^{2}\right) \int G\left( \vec{r},\omega \right) U\left( \vec{r}%
,\omega \right) d\vec{r}}
\end{eqnarray}%
The width of the conduction band is equal to $6A<<$ $\omega _{p_{0}}.$ To
carry out the integration in Eq. (\ref{36}), we consider the example of a
homogeneous medium with time dispersion. Then the operator $\hat{H}$ in (\ref%
{26}) is the Helmholtz operator 
\begin{equation}
\hat{H}=\Delta +\frac{\omega ^{2}}{c^{2}}\varepsilon (\omega ),  \label{38}
\end{equation}%
where $\varepsilon (\omega )$ is the dielectric permittivity of the host
medium, in which the periodic grating of small particles (local
perturbations) is located. For definiteness, we assume that the dielectric
constants of both host medium and particles are of the Drude form:%
\begin{equation}
\varepsilon _{0}\left( \omega \right) =1+\frac{\Omega ^{2}}{\omega
_{0}^{2}-\omega ^{2}};  \label{39}
\end{equation}%
\begin{equation}
\varepsilon \left( \omega \right) =1+\frac{\Omega _{1}^{2}}{\left( \omega
_{0}^{\prime }\right) ^{2}-\omega ^{2}}.  \label{40}
\end{equation}

\bigskip\ The dispersion equation Eq. (\ref{35}) has two solutions:

\begin{equation}
\omega _{_{1,2}}=\frac{\Omega _{1}^{2}+\left( \omega _{0}^{\prime }\right)
^{2}+\omega _{s}^{2}}{2}\pm \sqrt{\left( \frac{\Omega _{1}^{2}+\left( \omega
_{0}^{\prime }\right) ^{2}+\omega _{s}^{2}}{2}\right) ^{2}-\omega
_{s}^{2}\left( \omega _{0}^{\prime }\right) ^{2}}  \label{41}
\end{equation}%
where 
\begin{equation*}
\omega _{s}^{2}=\frac{4\pi c^{2}}{S}
\end{equation*}%
is the frequency of the ''geometrical'' resonance related to the finite
size, $a$, of the local perturbation (Debye resonance of the single
particle); $S=\int \frac{f(\vec{r})d\vec{r}}{r}\symbol{126}a^{2}$. \ Note
that the number of solutions depends on the explicit form of the functions $%
\varepsilon _{0}(\omega )$ and $\varepsilon (\omega )$, and can be larger
than two.

Multiple scattering between neighboring particles leads to the broadening of
the discrete levels $\omega _{_{1,2}}$ into passing zones of finite width.
Frequencies $\omega _{_{1,2}}=\omega _{_{01,02}}+\delta \omega _{_{1,2}}$
correspond to a peculiar kind of polaritons that arise in photonic crystals
with time dispersion due to the interaction of the electromagnetic field ($%
\omega $), eigenoscillations of the dielectric medium ($\omega _{0}),$ and
Debye resonance ($\omega _{s})$.

For the width of the interaction-induced passing zones, $\triangle \omega
_{1,2},$ Eq. (\ref{37}) yelds%
\begin{equation}
\triangle \omega _{1,2}\approx 6\frac{Ve^{-\left| \zeta _{m}\right| b}\omega
_{1,2}}{Sb},  \label{42}
\end{equation}

\begin{equation}
V=\int f\left( \vec{r}\right) d\vec{r}.  \label{43}
\end{equation}%
Terms independent on $\overrightarrow{k}$ (small shifts of $\omega _{_{1,2}}$%
) are omitted in Eq. (\ref{42}). Recall that for the exponent in the Green
function to be real (stability condition), it is nesessary that

\begin{equation}
1-\frac{\Omega ^{2}}{\omega _{1,2}^{2}}<0.  \label{basss}
\end{equation}%
One can see from (\ref{42}) that the zone width is maximal at $\omega
_{i}=\Omega $ and is equal $\Delta \omega _{i}=\frac{a}{b}\omega _{i}.$ For $%
\frac{\Omega ^{2}}{\omega _{0\text{ }1,2}^{2}}$ increasing, the width
decreases exponentially, and at $\frac{\Omega ^{2}}{\omega _{0\text{ }%
1,2}^{2}}>>1,$ the zone structure disappears. When the spacing between
levels,\ $\left| \omega _{1}-\omega _{2}\right| ,$ is of order of $\omega
_{s},$ the width of the passing zone is $\frac{b}{a}>>1$ times smaller than
the size of the band gap.

To conclude, time-spatial dispersion causes important differences in the
spectrum of photons in periodic dielectric systems from that of electrons in
solid state crystals. It gives rise to different types of waves, changes the
number and structure of passing zones and band gaps, and dramatically
deforms the Brillouin zones.

\end{document}